\documentclass[10pt]{article}
\usepackage{latexsym,amsmath,amsfonts,amssymb}
\newcommand{\RR}{\mbox{${\rm \:  R\!\!\!\! I
\;\;}$}}

\newtheorem{prop}{Proposition}

\newtheorem{thm}{Theorem}

\newtheorem{lem}{Lemma}

\def\qed{\relax\ifmmode\hskip2em \Box\else\unskip\nobreak\hskip1em $\Box$\fi}

\begin{document}
\centerline{\Large{\bf A note on Generalized Concurrences and
Entanglement Detection}}

\centerline{}

 \vspace{4ex}
\begin{center}
Laura Cattaneo and Domenico D'Alessandro\\
Department of Mathematics, Iowa State University, Ames IA 50011, U.S.A.\\
e-mail: cattaneo@iastate.edu, daless@iastate.edu
\end{center}

\begin{abstract}
We study   {\it generalized concurrences} as a tool to detect  the
entanglement of bipartite quantum systems. By considering the case
of $2\times 4$ states of rank 2, we prove that generalized
concurrences do not, in general,  give a necessary and sufficient
condition of separability. We identify a set of entangled states
which are undetected by this method.
\end{abstract}

\section{Introduction}

Consider a bipartite quantum system consisting of two subsystems $A$
and $B$, with associated Hilbert spaces ${\cal H}_A$ and ${\cal
H}_B$ of dimensions $n_A$ and $n_B$, respectively. The overall
system has Hilbert space ${\cal H}={\cal H}_A \otimes {\cal H}_B$ of
dimension $n:=n_An_B$. The state of the total system is represented
by an Hermitian $n \times n$ matrix $\rho$, called the {\it density
matrix}, which is positive semi-definite and has trace equal to one.
A density matrix is called {\it separable} if it can be written as
\begin{equation}\label{sepcond}
\rho=\sum_{j}\mu_j |\psi_{Aj}\rangle\langle \psi_{Aj}|  \otimes
|\psi_{Bj}\rangle\langle \psi_{Bj}|, \qquad \mu_j
>0, \qquad \sum_{j} \mu_j=1, \qquad |\psi_{Aj}\rangle (|\psi_{Bj}\rangle )\in {\cal
H}_A ({\cal H}_B).
\end{equation}
A state that is not separable is called {\it entangled}. One of the
fundamental open questions in quantum information theory is to give
criteria to decide whether a density matrix $\rho$ describing the
state of a bipartite quantum system represents an entangled or a
separable state.

Define the {\it partial transposition} of a $n_A n_B \times n_A n_B$
matrix $\rho=\sigma \otimes S$ (with $\sigma$ and $S$ of dimensions
$n_A \times n_A$ and $n_B \times n_B$, respectively) as
$\rho^{T_A}:=\sigma^T \otimes S$ and extend the definition to any
Hermitian matrix by linearity. A very popular test introduced in
\cite{HHH},\cite{Peres}, based on the partial transposition of
$\rho$, gives a criterion which is both simple and very powerful.
This test is called the {\it Positive Partial Transposition
(PPT)}-{\it test}.  It says that if $\rho$ is separable $\rho^{T_A}
\geq 0$. We shall call a state $\rho$ with $\rho^{T_A} \geq 0$  {\it
a $PPT$-state}. Therefore, every separable state is a PPT-state. The
converse has been proved to be true in the $2 \times 2$ and $2
\times 3$ cases \cite{HHH}, as well as in the $2\times N$ case with
rank lower than $N$ \cite{Kraus_etal}. The latter result has  been
generalized to $M\times N$ ($M<N$) and rank lower than $N$ in
\cite{HLVC}. On the other hand, higher dimensional examples have
been constructed of bipartite systems whose entanglement is not
detected by this test. This motivates the investigation of more
tests to detect entanglement in quantum systems.

Generalizing the definition of {\it concurrence} given by S. Hill
and W. Wootters \cite{HW}, \cite{Wootters} for the $2 \times 2$
case, A. Uhlmann introduced {\it generalized concurrences} in
\cite{Ul}. Generalized concurrences are functions of the state
$\rho$, $C_\Theta$, parametrized by a class of quantum symmetries
$\Theta$ (see next section for definitions and main properties).
 Separable states are such that all generalized concurrences are
equal to zero and A. Uhlmann proved that the converse is true for
the case of density matrices of rank 1 (pure states). He stated that
it is `unlikely' that this requirement can be dropped and we will
show in this paper that this is indeed the case. Generalized
concurrences give however an additional test of entanglement, that
is, if we can find a generalized concurrence, $C_\Theta$, such that
$C_\Theta(\rho)\not=0$, then $\rho$ is entangled. In this note, we
consider generalized concurrences in the simplest case  not
considered in \cite{Ul}, \cite{Wootters}. That is the case of $2
\times 4$ systems with density matrices of rank $2$. We shall see
that, as A. Uhlmann thought, even in this simple situation, the test
based on generalized concurrences is not necessary and sufficient
and there are entangled states that are undetected.

This paper is organized as follows. In Section \ref{badef} we give
the main definitions concerning generalized concurrences, describe
their role in entanglement detection and recall a connection between
certain quantum symmetries and Cartan involutions \cite{Helgason},
established in \cite{MikoBook}, \cite{mikofigata}, which we shall
use in the our derivation. Section \ref{mainres} presents the main
result. We consider bipartite systems where the two subsystems have
dimensions $2$ and $4$ respectively and assume that the density
matrix has rank $2$. Under these assumptions, the main result,
theorem \ref{UWtheo}, describes the class of density matrices for
which all the concurrences are zero. This set is made up of both
separable and entangled states. This shows that there are entangled
states that cannot be detected using generalized concurrences. Some
conclusions are drawn in section \ref{conclu}. Most of the technical
 proofs, including the proof of theorem \ref{UWtheo}, are presented in
 the Appendixes.

\section{Quantum symmetries, conjugations, generalized concurrences and Cartan
involutions} \label{badef} A {\it quantum symmetry} $\Theta$ is a
map $\Theta:$ ${\cal H} \rightarrow {\cal H}$, where ${\cal H}$ is
Hilbert space, of the form
$$
\Theta:=e^{i\phi} U, \qquad \phi \in \RR
$$
where $e^{i\phi}$ is a, physically irrelevant, phase factor and $U$
is either a unitary or anti-unitary operator. An anti-unitary
operator $U$ is defined by the two properties \begin{equation}
\nonumber \langle U \beta|U\alpha \rangle= \langle \beta | \alpha
\rangle^*,
\end{equation}
\begin{equation} \label{pol2}
 U \left(c_1| \alpha \rangle + c_2 |\beta \rangle \right)
=c_1^*U|\alpha \rangle+ c_2^* U |\beta \rangle,
\end{equation}
for each pair of vectors $|\alpha\rangle$ and $|\beta \rangle$ in
${\cal H}$ and pair of complex numbers $c_1$ and $c_2$. Property
(\ref{pol2}) is referred to as {\it anti-linearity}. A {\it (skew)
conjugation} is an anti-unitary quantum symmetry $\Theta$ satisfying
($\Theta^2=-{\bf 1}$)  $\Theta^2={\bf 1}$, where $\bf 1$ is the
identity operator.

Given a conjugation $\Theta$, A. Uhlmann \cite{Ul} defined a {\it
generalized concurrence} associated with $\Theta$ as a function on
${\cal H}$, $C_{\Theta}(|\psi\rangle)$, given by
\begin{equation}\label{genconcdef}
C_\Theta(|\psi\rangle):=|\langle \psi| \Theta |\psi \rangle|.
\end{equation}
This definition is extended to states represented by a general
density matrix $\rho$ using the {\it convex roof} procedure (cf.
\cite{Ul}). This means that $C_\Theta(\rho)$ is defined as
\begin{equation}\label{pol9}
C_\Theta(\rho):=\min \sum_j \mu_j C_{\Theta}(|\psi_j\rangle),
\end{equation}
where the minimum is taken among all the possible decompositions of
$\rho$ as \begin{equation}\label{decorho} \rho=\sum_j \mu_j |\psi_j
\rangle \langle \psi_j|, \qquad \mu_j >0, \qquad \sum_{j}\mu_j=1.
\end{equation}

Associated with a quantum symmetry  $\Theta$ is a super-operator
$\theta$, mapping linear Hermitian operators to linear Hermitian
operators, defined as
\begin{equation}\label{pop}
\theta(\rho):=\Theta \rho \Theta^{-1}.
\end{equation}
It can be easily verified that $\theta$ is linear, positive and
trace preserving map (in fact, $\theta (\rho)$ has the same
eigenvalues as $\rho$). While $\Theta$ determines $\theta$ according
to formula (\ref{pop}) specification of $\theta$ on all linear
Hermitian operators determines $\Theta$ up to a phase factor
\cite{GalPas}. If $\Theta$ is a conjugation or a skew-conjugation,
then $\Theta^{-1}=\pm \Theta$ and $\theta^2$ is equal to the
identity operator. A. Uhlmann \cite{Ul} gave a general method to
calculate generalized concurrences, that is to find the minimum in
(\ref{pol9}), in terms of the superoperator $\theta$ defined in
(\ref{pop}. We collect this result in the following theorem.
\begin{thm} \label{calcu} \cite{Ul} Assume $\Theta$ is a
conjugation and consider the matrix \begin{equation}\label{matcalcu}
\rho^{\frac{1}{2}} \theta(\rho)
\rho^{\frac{1}{2}}:=\rho^{\frac{1}{2}} \Theta \rho \Theta
\rho^{\frac{1}{2}}, \end{equation} which is positive semi-definite.
If $\lambda_{max}$ is its largest eigenvalue and $\lambda_1,\ldots,
\lambda_{n-1}$ are the remaining (possibly repeated) eigenvalues,
then
\begin{equation}\label{polop}
C_{\Theta} (\rho)= \max\{ 0, \sqrt{\lambda_{max}}-\sum_{j=1}^{n-1}
\sqrt{\lambda_j}\}.
\end{equation}
\end{thm}
\vspace{0.25cm}

In \cite{MikoBook}, \cite{mikofigata} a connection was recognized
between quantum symmetries and {\it Cartan involutions} used in
Cartan classification of symmetric spaces of the Lie group $SU(n)$.
This connection is useful to study the dynamics of generalized
concurrences as well as to parametrize conjugations and
skew-conjugations. We explain this next.

Consider a quantum symmetry $\Theta$ and the corresponding $\theta$
defined as in (\ref{pop}). Assume $\theta$ is such that $\theta^2$
is equal to the identity operator. Then $\Theta$ is a conjugation, a
skew-conjugation or a unitary symmetry whose square is a multiple of
the identity. We shall call a quantum symmetry with this property a
{\it Cartan involutory symmetry} for reasons that will be apparent
shortly.
 Consider the space of Hermitian operators on an $n$-dimensional
Hilbert space. As $u(n)$ denotes the space of $n \times n$
skew-Hermitian matrices, we denote by $i u(n)$ the space of $n\times
n$ Hermitian matrices. This space, when equipped with the
anti-commutator operation, $\{A, B\}:=AB+BA$, is a Jordan algebra
and $\theta$ defined in (\ref{pop}) is a Jordan algebra isomorphism
$iu(n) \rightarrow iu(n)$ satisfying $\theta^2={\bf 1}$, with ${\bf
1}$ the identity operator. Let ${\cal P}$ and ${\cal K}$ subspaces
of $u(n)$ such that $i{\cal P}$ and $i{\cal K}$ are the $+1$ and
$-1$, eigenspaces of $\theta$. The map $\tilde \theta$ on $u(n)$
defined by
\begin{equation}\label{invole}
\tilde \theta(A):=i \theta(iA).
\end{equation}
is a Lie algebra isomorphism of $u(n)$ and it is such that ${\cal
K}$ and ${\cal P}$ are the $+1$ and $-1$ eigenspaces of $\tilde
\theta$. Moreover $\tilde \theta^2={\bf 1}$. A Lie algebra
isomorphism with this property is called a {\it Cartan involution}
\cite{Helgason}. Therefore, there exists a one to one correspondence
given by formula (\ref{invole}) between Cartan involutions and
Cartan involutory symmetries. A Cartan involution $\tilde \theta$
induces a Cartan decomposition of the Lie algebra $u(n)$ which means
that the associated subspaces ${\cal K}$ and ${\cal P}$ satisfy the
commutation relations\footnote{The viceversa is also true. That is,
if one has a Cartan decomposition of the form (\ref{Cartan}) then
one can define a Cartan involution as a Lie algebra isomorphism
having ${\cal K}$ and ${\cal P}$ as the $+1$ and $-1$ eigenspaces,
respectively.}
\begin{equation}\label{Cartan}
[{\cal K}, {\cal K}] \subseteq {\cal K}, \qquad [{\cal K}, {\cal P}]
\subseteq {\cal P}, \qquad [{\cal P}, {\cal P}] \subseteq {\cal K}.
\end{equation}
 According to Cartan decomposition theorem \cite{Helgason}
 to the Lie algebra decomposition (\ref{Cartan}) there corresponds a
 decomposition of the Lie group $U(n)$, in that every element $X$ of
 $U(n)$ can be written as $X=KP$ where $K$ is in the connected Lie
 group, $e^{\cal K}$, associated with  ${\cal K}$ (which from the first one of
 (\ref{Cartan}) is a Lie subalgebra of $u(n)$) and $P$ is the
 exponential of an element in ${\cal P}$. The quotient space
 $U(n)/e^{\cal K}$ is called a {\it symmetric space} of $U(n)$.
 Cartan has classified all the symmetric spaces of $U(n)$ and
 therefore all the decompositions of the type (\ref{Cartan}) and all
 the Cartan involutions \cite{Helgason}. From the correspondence (\ref{invole}),
 this gives a classification of all the  Cartan involutory
 symmetries. There are three types of Cartan involutions labeled by
 {\bf AI}, {\bf AII} and {\bf AIII} which correspond, respectively
 to conjugations, skew-conjugations and unitary symmetries. In particular,
  it follows,
 using Cartan parametrization
  (cf. Tables 5.1 and 5.2 in \cite{MikoBook}), that up to a phase
  factor, every conjugation can be written as
  $\Theta_I(|\psi\rangle)=TT^T|\overline{ \psi} \rangle$ while every
  skew-conjugation can be written as
  \begin{equation}\label{ppp}
\Theta_{II} (|\psi\rangle)=TJT^T|\overline{ \psi}\rangle.
  \end{equation}
In these formulas and in the following $|\overline{ \psi} \rangle$ (
$\overline{\rho}$ ) denotes the complex conjugate of the vector
$|\psi \rangle$ (of a matrix $\rho$), $J$ is the matrix $J:=\begin{pmatrix}0 & {\bf 1} \\
-{\bf 1} & 0
\end{pmatrix}$, where ${\bf 1}$ is the $\frac{n}{2} \times
\frac{n}{2}$ (assuming $n$ even) identity and $T$ is an arbitrary
(parameter) special unitary matrix, i.e. a matrix in $SU(n)$. We now
turn to the application of this theory to entanglement.

\vspace{0.25cm}

When dealing with bipartite systems, whose Hilbert space ${\cal H}$
is the tensor product of two Hilbert spaces ${\cal H}:={\cal H}_A
\otimes {\cal H}_B$, it is natural to construct quantum symmetries
on ${\cal H}$ as tensor products of symmetries on ${\cal H}_A$ and
${\cal H}_B$.\footnote{The tensor product of two anti-linear
(linear) operators $\Theta_A \otimes \Theta_B$ is defined on product
vectors $|\psi_A\rangle  \otimes |\psi_B \rangle$ as $\Theta_A
\otimes \Theta_B (|\psi_A\rangle  \otimes |\psi_B
\rangle):=\Theta_A(|\psi_A\rangle) \otimes \Theta_B(|\psi_B\rangle)$
and then extended by anti-linearity (linearity) for linear
combinations of product vectors.} Consider two spaces ${\cal H}_A$
and ${\cal H}_B$ both with even dimension. A. Uhlmann \cite{Ul}
considers conjugations $\Theta$ constructed as tensor products of
skew-conjugations, $\Theta_A$ and $\Theta_B$. Using the
characterization of skew-conjugations (\ref{ppp}) it is
straightforward to prove the following.
\begin{thm}\label{zeroconc} If $\rho$ is a separable state, then
$C_\Theta(\rho)=0$ for every conjugation $\Theta=\Theta_A \otimes
\Theta_B$, with $\Theta_{A,B}$ skew-conjugations on ${\cal H}_{A,B}$
\end{thm}

\noindent{\bf Proof} We use  the general form of a skew-conjugation
(\ref{ppp}) and define $\Theta_{A,B}( |\psi_{A,B} \rangle ):=
T_{A,B} J T_{A,B}^T |\overline{ \psi_{A,B}} \rangle$, with general
matrices $T_A\in SU(n_A)$ and $T_B \in SU(n_B)$. Using the
decomposition (\ref{sepcond}) for $\rho$ with the definition
(\ref{pol9}) for the generalized concurrence $C_\Theta$,  we have
\begin{equation}
\nonumber  0 \leq C_{\Theta}(\rho) \leq \sum_j \mu_j
C_\Theta(|\psi_{Aj} \rangle \otimes |\psi_{Bj} \rangle)=0.
\end{equation}
The last equality is due to the fact that, with $\Theta:=\Theta_A
\otimes \Theta_B$, for each $j$, we have from (\ref{genconcdef}),
\begin{equation}\nonumber
C_\Theta(|\psi_{Aj} \rangle \otimes |\psi_{Bj} \rangle)=
\left|\langle \psi_{Aj}|\otimes \langle \psi_{Bj}| (\Theta_A \otimes
\Theta_B) |\psi_{Aj} \rangle \otimes |\psi_{Bj} \rangle \right|=
\end{equation}
$$
\left|\langle \psi_{Aj}|\otimes \langle \psi_{Bj}|T_A J T_A^T
\otimes T_B J T_B^T  | \overline{ \psi_{Aj}} \rangle \otimes |
\overline{ \psi_{Bj}} \rangle \right|= \left| \langle T^\dagger_A
\psi_{Aj} | J | \overline{T_A^\dagger \psi_{Aj}}\rangle \right|
\times \left| \langle T^\dagger_B \psi_{Bj} | J | \overline{
{T_B^\dagger \psi_{Bj}}}\rangle\right|=0,
$$
as both factors in the last expression are zero. \qed

This theorem gives a method to detect  entanglement. If there exists
a $\Theta$ such that $C_{\Theta}(\rho) >0$, then $\rho$ is
entangled. To calculate $C_\Theta$ one can use formula
(\ref{polop}). If $C_{\Theta}(\rho)=0$ for every $\Theta$, then it
may be separable or entangled (as we shall see below) and, in the
latter case, entanglement is not detected by this method.

In the special case of two qubits, i.e., $n_A=n_B=2$,  $T_A$ and
$T_B$ are general matrices in $SU(2)$ and, for every matrix $T$ in
$SU(2)$, \begin{equation}\label{opla} TJT^T=J.
\end{equation} As a
consequence of formula (\ref{ppp}) there is only one (generalized)
concurrence corresponding to the conjugation $\Theta (|\psi
\rangle)=J \otimes J |\overline{ \psi} \rangle$. This is the
concurrence originally considered by Hill and Wootters in \cite{HW},
\cite{Wootters}. In this case the converse of theorem \ref{zeroconc}
holds, that is, if the (only) concurrence is zero the state is
separable. It was proven in \cite{Ul} that the converse also holds
when $\rho$ is a pure state. In the next section, we settle in the
negative the question of whether the converse holds in general.


\section{Main Result}

\label{mainres} From now on, we shall consider only, the case of a
state $\rho$ on a Hilbert space ${\cal H}_A \otimes {\cal H}_B$ with
$\dim{\cal H}_A:=n_A=2$ and $\dim{\cal H}_B:=n_B=4$ and $\rho$ of
rank $2$, although some of things we shall say can be extended
without difficulties to the general case. In the following we shall
also denote by $J_{2m}$ the $ 2m \times 2m $ matrix
$J_{2m}=\left(\begin{array}{cc}0 & {\bf 1}  \\ -{\bf 1} & 0
\end{array}\right)$ where ${\bf 1}$ is the $m \times m$ identity. If
$\Theta$ is the tensor product of two skew-conjugations, then, using
(\ref{ppp}), $\Theta$ has the form $\Theta |\psi \rangle =T_A J_2
T_A^T \otimes T_B J_4 T_B^T |\overline{ \psi} \rangle$ with $T_A \in
SU(2)$ and $T_B \in SU(4)$. From the above recalled property
(\ref{opla}) of $T_A \in SU(2)$ we have that defining
\begin{equation}\label{emme} M:=J_2\otimes TJ_4T^T,
\end{equation}
with $T \in SU(4)$, every conjugation $\Theta$ which is tensor
product of two skew-conjugations can be written as
\begin{equation} \label{thmo} \Theta (|\psi \rangle)=M |\overline{
\psi}\rangle
\end{equation} and by varying $T \in SU(4)$ we obtain all of such
products. A state $\rho$ which has zero concurrence $C_\Theta$ for
all such $\Theta$'s will be called a {\it Zero
Concurrence($ZC$)-state}. Therefore from theorem \ref{zeroconc} it
follows that separable states are $ZC$-states. In theorem
\ref{UWtheo}, we shall see that the set of $ZC$-states is made up of
two nonempty subsets containing respectively only separable and
 only entangled states.

\vspace{0.25cm}

All the properties of $\rho$ which are of interest to us
(separability, $PPT$ and $ZC$) are invariant under local
transformations, i.e., under transformations of the form $\rho
\rightarrow (X_1 \otimes X_2) \rho (X_1^\dagger  \otimes
X_2^\dagger)$ with $X_1 \in SU(2)$ and $X_2 \in SU(4)$. Namely, we
have the following.

\begin{prop}\label{loceq}
For every $X_1 \in SU(2)$ and $X_2 \in SU(4)$
\begin{enumerate}
\item $\rho$ is separable if and only if $(X_1 \otimes X_2) \rho (X_1^\dagger \otimes
X_2^\dagger)$ is separable.
\item $\rho$ is $PPT$ if and only if $(X_1 \otimes X_2) \rho (X_1^\dagger \otimes
X_2^\dagger)$ is $PPT$.
\item $\rho$ is $ZC$ if and only if $(X_1 \otimes X_2) \rho (X_1^\dagger \otimes
X_2^\dagger)$ is $ZC$.
\end{enumerate}
\end{prop}
{\bf Proof.} The first two properties are obvious. Now, assume that
$\rho$ is a $ZC$-state and let $\Theta$ a general conjugation
(\ref{thmo})  corresponding to a matrix $M$ as in  (\ref{emme}). Let
$\tilde \Theta$
 a conjugation corresponding to matrix
 $\tilde M:=J_2 \otimes X_4^\dagger T J_4 T^T \overline{ X_4}$. Since $\rho$ is $ZC$,
 $C_{\tilde \Theta}(\rho)=0$. In particular, there exists a decomposition
 of $\rho$ as in (\ref{decorho}) such that, for every $j$,
 \begin{equation}\label{iol}
0=C_{\tilde \Theta}(|\psi_j \rangle)= \left| \langle \psi_j |J_2
\otimes X_4^\dagger T J_4 T^T \overline{ X_4} |\overline{ \psi_j}
\rangle \right|= \left| \langle \psi_j |X_2^\dagger \otimes
X_4^\dagger (J_2 \otimes T J_4 T^T )\overline{ X_2} \otimes
\overline{ X_4} |\overline{ \psi_j} \rangle \right|.
\end{equation}
However the last term of (\ref{iol}) is $C_{\Theta}(X_2 \otimes X_4
|\psi_j \rangle) $ and $\{ \mu_j, X_2 \otimes X_4 |\psi_j \rangle
\}$ give a decomposition of $(X_2 \otimes X_4) \rho (X_2^\dagger
\otimes X_4^\dagger)$. Since $\Theta$ is arbitrary $(X_2 \otimes
X_4) \rho (X_2^\dagger \otimes X_4 \dagger)$ is $ZC$ as well. The
converse follows immediately from the fact that $X_2$ and $X_4$ are
arbitrary. \qed

\vspace{0.25cm}

The previous property suggests to place $\rho$ is in a {\it
canonical form} using only local transformations, without loss of
generality. We shall describe this canonical form next. Since $\rho$
has rank $2$, we write it in terms of its  eigenvectors
corresponding to nonzero eigenvalues as
\begin{equation} \label{rhok} \rho=\lambda |\psi_1 \rangle \langle \psi_{1}|
+(1-\lambda)|\psi_2 \rangle \langle \psi_2| \,,
\end{equation}  with
$0 < \lambda <1$. We assume that at least one between $|\psi_1
\rangle$ and $|\psi_2 \rangle$ is an entangled pure
state.\footnote{There are several general methods to check that a
pure bipartite state is entangled. An example is given by the
entropy cf., e.g., \cite{NC}.} If both are $|\psi_1\rangle$ and
$|\psi_2\rangle$ are separable, then $\rho$ is also  separable and
therefore it is $ZC$ and $PPT$. Excluding this case, we assume
$|\psi_1\rangle$ entangled. Using Schmidt decomposition theorem (cf.
Theorem 2.7 in \cite{NC}) we  choose orthonormal bases $\{ |a_{1,2}
\rangle \}$ of $ {\cal H}_A$ and $\{ |b_{1,2,3,4}\rangle \}$ of
${\cal H}_B$ such that $|\psi_1\rangle =q_1 |a_1,b_1 \rangle + q_6
|a_2,b_2 \rangle$, with both $q_1$ and $q_6$ real and nonnegative.
Moreover since $|\psi_1 \rangle$ is entangled, both $q_1$ and $q_6$
are strictly positive. Using these bases, we write $|\psi_2
\rangle:=\sum_{j=1,2, \, k=1,\ldots,4} r_{jk} |a_j,b_k\rangle$. We
use a local transformation of the form ${\bf 1} \otimes X$ where
$X\in SU(4)$ acts as the identity on the subspace spanned by $|b_1
\rangle$ and $|b_2 \rangle$, to set the coefficient $r_{14}$ to zero
without changing $|psi_1 \rangle$. Finally, since $|\psi_2 \rangle$
(and $|\psi_1\rangle$) is defined up to an overall phase factor we
assume $r_{11}$ real and nonnegative. Since $\langle \psi_1 |\psi_2
\rangle=0=q_1 r_{11}+q_6r_{22}$ which forces $r_{22}$ to be real and
non-positive. In conclusion, $\rho$ in (\ref{rhok}) is such that
either both $|\psi_1\rangle$ and $|\psi_2\rangle$ are separable (and
it is therefore separable) or it can be transformed with local
transformations into a a canonical form where the $|\psi_1\rangle$
and $|\psi_2\rangle$ coordinates are, respectively,
$\psi_1:=(q_1,0,0,0,0,q_6,0,0)^T $,
$\psi_2:=(p_1,p_2,p_3,p_4,p_5,p_6,p_7,p_8)^T$ with $q_1>0$ and
$q_6>0$, $p_1 \geq 0$, $p_6 \leq 0$, $p_4=0$ and $p_1q_1+p_6q_6=0$.
This is the canonical form we shall refer to in the sequel.

\vspace{0.25cm}

We now state  a fact  which is a special case of a general result
proven in  \cite{HLVC}, \cite{Kraus_etal}.
\begin{prop}\label{KrausB}
Assume $\rho$ is a  $2 \times 4$ state with rank $2$. Then $\rho$ is
separable if and only if it is $PPT$.
\end{prop}
The above proposition says that the PPT test characterizes
completely separable and entangled states in the $2 \times 4$, rank
$2$, case. Some more information we shall use is given in the
following Lemma.
\begin{lem}\label{moreinfo}
Let $\rho$ be a state in canonical form. Then $\rho$ is  $PPT$ and
therefore separable if and only if it has the form
\begin{equation}\label{formPPT}
\rho=\left(\begin{array}{cccc}\rho_{11} & 0 & \rho_{12} & 0 \\ 0 & 0
& 0 & 0\\ {\rho_{12}}^\dag & 0 & \rho_{22} & 0 \\ 0 & 0 & 0
&0\end{array}\right),
\end{equation}
where the $4\times 4$ matrix
\begin{equation}
\tilde{\rho}:=\left(\begin{array}{cc}\rho_{11} & \rho_{12} \\
{\rho_{12}}^\dag & \rho_{22}\end{array}\right) \label{rhotilde}
\end{equation}
is separable as a two qubit state.
\end{lem}
We give the proof of this Lemma in Appendix C. This also gives an
alternative proof of Proposition \ref{KrausB}.

\vspace{0.25cm}

We are now ready to state our main result which describes completely
the set of, $2 \times 4$,  $ZC$-states of rank two. The proof is
given in Appendix B while auxiliary results are presented in
Appendix A.

\begin{thm} \label{UWtheo} A $2 \times 4$, rank $2$, state $\rho$ is a $ZC$-state if
and only if it is in one of the following two disjoint classes.
\begin{itemize}
\item The class $ZCS$  ($S$
stands for separable) which is defined as containing  states of the
form (\ref{rhok}) with $|\psi_1 \rangle$ and $|\psi_2 \rangle$
separable along with $PPT$-states which can be written in  canonical
form as in (\ref{formPPT}).
\item States of the form (\ref{rhok}) which can be written in
canonical form with
\begin{equation}\label{ZCEstati}
\lambda=\frac{1}{2}, \qquad  |\psi_1 \rangle= q_1 |a_1,b_1 \rangle+
q_6|a_2,b_2 \rangle, \qquad  |\psi_2\rangle=q_1 |a_1,b_3 \rangle +
q_6e^{i \phi} |a_2,b_4 \rangle \qquad \phi \in \mathbb{R}.
\end{equation}
These
states will be called $ZCE$-states ($E$ stands for entangled).
\end{itemize}
\end{thm}

\section{Conclusions}
\label{conclu}

Several extensions of the concurrence originally defined by Hill and
Wootters \cite{HW}, \cite{Wootters} for the $2 \times 2$ case  have
been proposed in the literature (see e.g., \cite{Horoall}). In few
cases  a direct physical in terms of probability for  the
measurements of appropriate observables has been indicated
\cite{Cirone}. However in most cases, a direct physical
interpretation is missing. The generalized concurrences considered
here  are the ones studied in \cite{Ul}. They are functions
constructed through anti-linear operators (symmetries). As observed
in \cite{Ul} these operators are intrinsically non-local as there is
no way to tensor them with the identity, that is, to apply them to a
part of the system by leaving the other unchanged. Using these
operators, a family of functions can be constructed which are all
zero if the state is separable. The question then arises on whether
these functions provide a complete test to detect entanglement. In
this note, we have given a negative answer to this question. The PPT
test is necessary and sufficient for entanglement of $2 \times 4$
states of rank $2$ \cite{HLVC}, \cite{Kraus_etal}. Generalized
concurrences can be used to detect entanglement, but in this case
they do not detect entanglement for a class of states ($ZCE$ states)
we have described in theorem \ref{UWtheo}.

In spite of this negative result, we believe  generalized
concurrences are still worth further investigation. In particular,
it is an open question whether for higher dimensional problems,
and-or higher rank, generalized concurrences may detect entanglement
of PPT states.

\vspace{0.25cm}

\noindent {\bf Acknowledgment.}

This work was supported by NSF, under CAREER grant ECS0237925. The
authors would like to thank Anna Sanpera for bringing references
\cite{HLVC} and \cite{Kraus_etal} to their attention.

\subsection*{Appendix A: Two Auxiliary Lemmas}\label{auxil}

The matrix $M$ in \eqref{emme} determines the particular generalized
concurrence considered. $ZC$-states, by definition, have all the
concurrences equal to zero. In principle $M$ depends on $15$
parameters since it depends on the matrix  $T$, which is a general
matrix in $SU(4)$ whose dimension is $15$. However, the form of $M$
can be greatly simplified. Using the Cartan decomposition of type
{\bf AII} \cite{Helgason}, every $T\in SU(4)$ can be written as
$T=PK$, where $K$ is symplectic and $P=e^G$ with
$G\in\mathbf{sp}(2)^{\bot}$. Matrices in
$\mathbf{sp}(2)^{\bot}$\footnote{The Lie algebra $\mathbf{sp}(m)$ is
defined as the one of skew-Hermitian $2m \times 2m$ matrices $A$,
satisfying $AJ_{2m}+J_{2m}A^T=0$. The orthogonal complement
$\mathbf{sp}(m)^{\bot}$ in $u(2m)$ is taken with respect to the
inner product $(A,B):=Trace(AB^\dagger)$.} have the form
\begin{equation}G=\left(\begin{array}{cc} A & bJ_2 \\
\overline{ b} J_2 & A^T \end{array}\right), \label{G_sp2T}
\end{equation}
with $A$ $2 \times 2$ skew-Hermitian and $b$ a complex scalar. Since
every symplectic matrix $K$ is by definition such that
$KJ_4K^T=J_4$, we can rewrite every $M$ in (\ref{emme}) in the form
\begin{equation}\label{emmeimpo}
M=J_2 \otimes e^{Gt} J_4 e^{ G^T t }, \qquad t \in \RR.
\end{equation}
Defining $H:=2GJ_4$ and $\eta:=\frac{1}{2}\sqrt{Tr(HH^\dagger)}$,
the following relations are easily verified:
\begin{equation}\label{relationsM}
GJ_4=J_4G^T:=\frac{1}{2} H, \qquad GH+HG^T=-\eta^2 J_4.
\end{equation}
The first lemma of this appendix gives a simplified expression for
$M$.
\begin{lem}\label{emmelemma} For any $M$ in (\ref{emme}),
$M \not=J_2 \otimes J_4$, there exists a $G$ (\ref{G_sp2T}) and
$H=2GJ_4$ and $\eta=\frac{1}{2} \sqrt{Tr(HH^\dagger)} \not=0$ such
that
\begin{equation}\label{emmesimplif}
M=J_2 \otimes \left( \cos(\eta t) J_4 + \frac{\sin (\eta t)}{\eta}
H\right).
\end{equation}
\end{lem}
{\bf Proof.} If $M \not= J_2 \otimes J_4$ then $G\not=0$ and
therefore $H \not=0$ and $\eta\not=0$.  From \eqref{emmeimpo}, it is
sufficient to prove that
$$
F_1(t):=e^{Gt}J_4 e^{G^T t}=\cos(\eta t) J_4 + \frac{\sin (\eta
t)}{\eta} H=:F_2(t).
$$
The matrix functions $F_1$ and $F_2$ are such that $\dot
F_1=GF_1+F_1G^T$ and $\dot F_2=GF_2+F_2G^T$. The first equation is
straightforward, while the second one follows from the relations in
\eqref{relationsM}. Since $F_1$ and $F_2$ satisfy the same
differential equations and are equal at $t=0$ they are the same for
every $t$.\qed\\

\noindent {\bf Remark.}\ For $\eta=0$, $H$ and $G$ are equal to zero
and $M$ becomes
\begin{equation}\label{emme0}
M=J_2 \otimes J_4.
\end{equation}
This expression can be obtained as the limit of (\ref{emmesimplif})
when
$\eta \rightarrow 0$. \qed\\

In the next result, we consider a general symmetric $2
\times 2$ complex matrix $C=\left(\begin{array}{cc} \alpha & \beta\\
\beta & \gamma\end{array}\right)$ and a diagonal matrix
$\Lambda=\left(\begin{array}{cc} \lambda & 0\\ 0 & 1-\lambda
\end{array}\right)$, with $0 < \lambda < 1$. We are interested in the eigenvalues
of the positive semidefinite matrix $B:=\sqrt{\Lambda}C\Lambda
C^\dag\sqrt{\Lambda}$, $\lambda_{max}$ and $\lambda_{min}$,  and, in
particular, in whether or not they are equal. The following lemma
gives necessary and sufficient conditions for this to happen.
\begin{lem}
The two eigenvalues of $B$ defined above, $\lambda_{max}$ and
$\lambda_{min}$, are equal if and only if the following two
conditions are verified.
\begin{itemize}
\item [(i)] $\lambda|\alpha|=(1-\lambda)|\gamma|$;
\item [(ii)] $\alpha \gamma \overline{ \beta}^2 \leq 0.$
\end{itemize}
\label{c1}
\end{lem}
\noindent{\bf Proof.} The eigenvalues $\lambda_{max}$ and
$\lambda_{min}$ are equal if and only if
$(\lambda_{max}-\lambda_{min})^2=(\operatorname{Tr}(B))^2-4\det
B=0$. Using the explicit expression of $B$,
\[B=\left(\begin{array}{cc} \lambda^2|\alpha|^2+\lambda(1-\lambda)|\beta|^2 &
\sqrt{\lambda(1-\lambda)}(\lambda\alpha\overline{\beta}+(1-\lambda)\beta\overline{\gamma})\\
\sqrt{\lambda(1-\lambda)}(\lambda\overline{\alpha}\beta+(1-\lambda)
\overline{\beta}\gamma) &
\lambda(1-\lambda)|\beta|^2+(1-\lambda)^2|\gamma|^2\end{array}\right),\]
we calculate \begin{eqnarray*} (\operatorname{Tr}(B))^2-4\det B &=&
(\lambda^2|\alpha|^2+2\lambda(1-\lambda)|\beta|^2
+(1-\lambda)^2|\gamma|^2+2\lambda(1-\lambda)|\alpha\gamma-\beta^2|)\\
&&\cdot(\lambda^2|\alpha|^2+2\lambda(1-\lambda)
|\beta|^2+(1-\lambda)^2|\gamma|^2-2\lambda(1-\lambda)|\alpha\gamma-\beta^2|)\,.
\end{eqnarray*}
The first factor in this expression is zero only if
$\alpha=\beta=\gamma=0$. If this is not the case,  we must have
\begin{equation}\label{oi}
\lambda^2|\alpha|^2+2\lambda(1-\lambda)|\beta|^2+
(1-\lambda)^2|\gamma|^2=2\lambda(1-\lambda)|\alpha\gamma-\beta^2|.
\end{equation}
Since this equation is trivially verified also in the special case
$\alpha=\beta=\gamma=0$, it is necessary and sufficient to have
$\lambda_{max}=\lambda_{min}$. Equation (\ref{oi}) can be written in
the simpler form $(i)$ and $(ii)$ proceeding as follows.

By the triangular inequality, we have that
\[2\lambda(1-\lambda)|\alpha\gamma-\beta^2|\leqslant
2\lambda(1-\lambda)(|\alpha\gamma|+|\beta|^2)\] and thus
\[\lambda^2|\alpha|^2+(1-\lambda)^2|\gamma|^2-
2\lambda(1-\lambda)|\alpha\gamma|\leqslant 0\,.\] But the l.h.s. of
the last inequality is equal to
$(\lambda|\alpha|-(1-\lambda)|\gamma|)^2$ and thus it is positive.
Hence, $\lambda|\alpha|-(1-\lambda)|\gamma|=0$, i.e., $ (i)$ is
satisfied. If we insert this condition in \eqref{oi}, we get
\[\lambda^2|\alpha|^2+\lambda(1-\lambda)|\beta|^2=
\lambda(1-\lambda)|\alpha\gamma-\beta^2|\,,\] where
$\lambda^2|\alpha|^2$ can be rewritten as
$\lambda(1-\lambda)|\alpha\gamma|$, because of (i). We then divide
both sides of the equation by $\lambda(1-\lambda)$ (since
$0<\lambda<1$, we have $\lambda(1-\lambda)\neq 0$). We obtain
$|\alpha\gamma|+|\beta|^2=|\alpha\gamma-\beta^2|$,
which is equivalent to  condition $(ii)$.\\

Conversely, if conditions (i) and (ii) are satisfied, then
\[\lambda^2|\alpha|^2+2\lambda(1-\lambda)|\beta|^2+
(1-\lambda)^2|\gamma|^2-2\lambda(1-\lambda)|\alpha\gamma-\beta^2|=
2\lambda(1-\lambda)(|\alpha\gamma|+|\beta|^2-|\alpha\gamma-\beta^2|)=0\,,\]
i.e., equation (\ref{oi}).\qed

\section*{Appendix B: Proof of theorem \ref{UWtheo}} \label{UW_proof}

The fact that states of the form (\ref{rhok}) with $|\psi_1 \rangle$
and $|\psi_2\rangle$ both separable are $ZC$ and  separable follows
from theorem \ref{zeroconc}. Let us therefore consider a state in
canonical form without loss of generality (cf. Proposition
\ref{loceq}). A state is a ZC-state if and only if the matrix
$\rho^{\frac{1}{2}} \theta(\rho) \rho^{\frac{1}{2}}$ in
\eqref{matcalcu} of theorem \ref{calcu}  has two coinciding
eigenvalues, for every $\Theta$ in (\ref{thmo}) with $M$ in
(\ref{emme}), (\ref{emmesimplif}). By writing $\rho$ as $U \tilde
\Lambda U^\dagger$, with $U$ unitary and $\tilde \Lambda$ equal to
zero except for the first two entries on the diagonal which are
equal to $\lambda$ and $1-\lambda$, it can be  seen that the
eigenvalues of $\rho^{\frac{1}{2}}\theta(\rho) \rho^{\frac{1}{2}}$
are the same as the eigenvalues of a $2 \times 2$ matrix of the form
$B$ considered in Lemma \ref{c1}.\footnote{We have (cf.
(\ref{thmo}), (\ref{pop})) $\rho^{\frac{1}{2}} \theta(\rho)
\rho^{\frac{1}{2}}= \rho^{\frac{1}{2}}M \overline{\rho} M^\dagger
\rho^{\frac{1}{2}}$, and, therefore, $\rho^{\frac{1}{2}}
\theta(\rho)
\rho^{\frac{1}{2}}=U\tilde{\Lambda}^{\frac{1}{2}}U^\dagger
M\overline{U} \tilde{\Lambda} U^TM^\dagger U
\tilde{\Lambda}^{\frac{1}{2}} U^\dagger$. If we denote by  $\tilde
C$ the symmetric matrix $U^\dagger M \overline{U}$, the eigenvalues
of $\rho^{\frac{1}{2}} \theta(\rho) \rho^{\frac{1}{2}}$ are the same
as the eigenvalues of $\tilde \Lambda^{\frac{1}{2}} \tilde C \tilde
\Lambda \tilde C^\dagger \tilde \Lambda^{\frac{1}{2}}$. Calling
$\Lambda$ ($C$) the upper $2\times 2$ block of the matrix $\tilde
\Lambda$ ($\tilde C$), since all the other entries of $\tilde
\Lambda$ are zeros, it follows that the nonzero eigenvalues are the
same as the ones of a matrix of the form $B$ considered in Lemma
\ref{c1}.} In this case $\lambda$ and $1-\lambda$ are the
eigenvalues of $\rho$ as in (\ref{rhok}) and $\alpha=\langle \psi_1|
M |\overline{ \psi_1}\rangle$, $\beta=\langle \psi_1| M |\overline{
\psi_2} \rangle $, $\gamma= \langle \psi_2| M | \overline{ \psi_2}
\rangle $, with $|\psi_1 \rangle $ and $|\psi_2 \rangle $ also as in
\eqref{rhok} and for every $M$ in \eqref{emme}. In the following
discussion we shall always assume, without loss of generality, that
the state $\rho$ is in canonical form.

If we calculate the explicit form for $\alpha$ and $\gamma$, using
the expression for $M$ in \eqref{emmesimplif}, \eqref{emme0},
\eqref{G_sp2T}, \eqref{relationsM}, we obtain
\begin{equation}\label{alfa}
\alpha=-4b \frac{\sin{\eta t}}{\eta} q_1 q_6,
\end{equation}
\begin{equation}\label{gamma}
\gamma=4 \frac{\sin(\eta t)}{\eta} \left( \overline{ b} \overline{
w}_2^T J_2 \overline {w}_4-b \overline{ w}_1^T J_2 \overline{ w}_3
\right) + 2 \operatorname{Tr}\left[ \left( \cos(\eta t) {\bf 1} + 2
\frac{\sin(\eta t)}{\eta} A \right) \left( \overline{ w}_4
\overline{ w}_1^T - \overline{ w}_2 \overline{ w}_3^T \right)
\right],
\end{equation}
where we have partitioned $|\psi_2 \rangle $ as $ |\psi_2 \rangle
:=(w_1^T, w_2^T,w_3^T,w_4^T)^T$ for $2$-dimensional vectors $w_j$,
$j=1,\ldots,4$. If $\rho$ is a $ZC$-state,  equation $(i)$ of Lemma
\ref{c1} has to hold with $\alpha$ and $\gamma$ for every
skew-Hermitian zero trace matrix $A$, every real $t$, and every
complex number $b$. In particular, by setting $b=0$ and varying $t$
and $A$, we obtain that it must be \begin{equation}\label{poi} w_4
w_1^T = w_2 w_3^T,
\end{equation}
and the second term in the r.h.s. of \eqref{gamma} is zero.
Inserting this constraint in $(i)$ of Lemma \ref{c1}, we have that
for every complex number $b$
$$\frac{\lambda}{1-\lambda}|b| q_1 q_6=|b w_2^T J_2 w_4-\overline{
b}
w_1^TJ_2 w_3|$$ must hold. For this to be verified one and only one
between $w_2^T J_2 w_4$ and $w_1^TJ_2 w_3$ must be different from
zero and equal to $\frac{\lambda}{1 - \lambda} q_1 q_6$ in absolute
value. Let us indicate by $ZCS$, $ZC$-states such that $w_1^TJ_2
w_3\not=0$ and by $ZCE$ its complement in the set of $ZC$-states. If
a state is $ZCS$, multiplying (\ref{poi}) on the right by $J_2w_1$
and using the fact that $w_3^TJ_2 w_1 \neq0$ but $w_1^TJ_2 w_1 =0$,
we obtain $w_2=0$. Analogously, multiplying by $J_2 w_3$ we obtain
$w_4=0$. In a similar fashion for $ZCE$ states, we obtain $w_1=0$
and $w_3=0$. Summarizing, if a state is $ZC$, it has to be of the
form $ZCS$ with
\begin{equation} \label{ZCSmini}
w_2=w_4=0, \qquad |w^T_1 J_2 w_3|=\frac{\lambda}{1-\lambda} q_1 q_6,
\end{equation}
or of the form $ZCE$ with
\begin{equation} \label{ZCEmini}
w_1=w_3=0, \qquad |w^T_2 J_2 w_4|=\frac{\lambda}{1-\lambda} q_1 q_6,
\end{equation}

In order to analyze the implications of the condition $(ii)$ of
Lemma \ref{c1}, we write $\beta$ in the two cases, $ZCS$ and $ZCE$,
and denote it by $\beta_S$ and $\beta_E$, respectively. With
$v_1=(q_1,0)^T$ and $v_2=(0, q_6)^T$, we obtain
\begin{equation}\label{betaS}
\beta_S=2b\frac{\sin(\eta t)}{\eta} \left( - v_1^TJ_2 \overline{
w}_3+v_2^TJ_2 \overline{ w}_1 \right),
\end{equation}
\begin{equation}\label{betaE}
\beta_{E}=v_1^T \left( \cos(\eta t) {\bf 1} +2 \frac{\sin(\eta t)
A}{\eta} \right) \overline{ w}_4 - v^T_2 \left( \cos(\eta t) {\bf 1}
+2 \frac{\sin(\eta t) A}{\eta} \right) \overline{ w}_2.
\end{equation}
Let us consider the case of $ZCE$-states first. Inserting
(\ref{ZCEmini}) and (\ref{poi}) in (\ref{gamma}) and using $\beta_E$
in (\ref{betaE}) for $\beta$, we obtain from condition $(ii)$
\begin{equation}
-16 q_1q_6 |b|^2\frac{\sin^2(\eta t)}{\eta^2} \overline{w}_2^T J_2
\overline{w}_4 \left[ v_1^T \left( \cos(\eta t) {\bf 1} +2
\frac{\sin(\eta t) \overline{A}}{\eta} \right) w_4 - v^T_2 \left(
\cos(\eta t) {\bf 1} +2 \frac{\sin(\eta t) \overline{ A}}{\eta}
\right) w_2 \right]^2\leq 0\,.\label{kl}
\end{equation}
This expression has to hold for every skew-Hermitian matrix $A$,
every $t$, and every $\eta \neq0$. Setting $A=0$ and recalling the
definition of the $v_{1,2}$ and $w_{2,4}$ vectors, and the fact that
$p_4=0$, we obtain $\overline {p}_3 \overline{p}_8 p_7^2 \geq 0$.
Setting $A=\left(\begin{array}{cc} i & 0 \\ 0 &
-i\end{array}\right)$ and $\cos(\eta t)=0$, we obtain
$-\overline{p}_3 \overline{p}_8 p_7^2 \geq 0$, which shows that
$p_7=0$, since $ p_3  p_8 = w_2^T J_2 w_4 \not=0$. Using this to
simplify (\ref{kl}), we find that $p_3$ and $p_8$ must be such that,
for every complex number $c$
$$ \overline{p}_3 \overline{p}_8 \left( c q_1 p_8 +\overline{c}q_6 p_3 \right)^2
\geq 0.
$$
It is easily seen that this is the case if and only if $q_1|
p_8|=q_6 |p_3|$. Combining this with \eqref{ZCEmini} and the fact
that $\| |\psi_2\rangle \|=1$, we find that we must have
$\lambda=\frac{1}{2}$ and $|p_8|=q_6$, $|p_3|=q_1$. Hence, states of
the type $ZCE$ must be of the form (\ref{ZCEstati}).\footnote{Notice
that a straightforward application of the PPT criterion shows that
these states are entangled.}
 Consider  $ZCS$-states next. In this case,  using (\ref{ZCSmini}),
 (\ref{alfa}), (\ref{gamma}) and (\ref{betaS}) with $b\not=0$, we
 obtain that condition $(ii)$ of Lemma \ref{c1} gives $\overline{w}_1
 ^TJ_2 \overline{w}_3 \left( - v_1^TJ_2 w_3 + v_2^T J_2 w_1
 \right)^2 \leq 0$. Writing this in terms of the vectors $|\psi_1 \rangle $
 and  $|\psi_2 \rangle $, we obtain the condition
 \begin{equation}\label{specialcon}
\left( - \overline{p}_2 \overline{p}_5 + \overline{p}_1
\overline{p}_6 \right) \left( -q_1 p_6 -q_6p_1 \right)^2 \leq 0,
 \end{equation}
which supplements (\ref{ZCSmini}) in describing these states. To
show that these states correspond to the ones in (\ref{formPPT}), we
consider the two qubit state
\[\tilde\rho=\lambda |\tilde \psi_{1}\rangle \langle \tilde \psi_1| +
(1-\lambda) |\tilde \psi_2\rangle \langle  \tilde \psi_2| \] with
\[ |\tilde\psi_1\rangle =(q_1, 0, 0, q_6)^T, \quad |\tilde \psi_2 \rangle =(p_1, p_2, p_5,
p_6)^T,\] corresponding to \eqref{rhotilde}. We have to show that
that $\tilde{\rho}$ is separable. For this we use the \emph{two
qubit} concurrence \cite{Wootters} which gives a necessary and
sufficient condition of separability. There is only one concurrence
in the two qubit case, which can be defined as in (\ref{polop}),
where $\lambda_{max}$, $\lambda_{1,2,3}$ are the eigenvalues of the
matrix
\[\tilde \rho^{\frac{1}{2}} J_2 \otimes J_2 \overline{\tilde \rho}
J_2\otimes J_2 \tilde \rho^{\frac{1}{2}}\,.\] A two qubit state
$\tilde \rho$ is separable if and only if the concurrence is zero.
Using the fact that the state has  rank two and proceeding as for
the $2 \times 4$ case, now with $M=J_2 \otimes J_2$, we have that
this is verified if and only if both conditions of Lemma  \ref{c1}
are verified, with $\alpha$, $\beta$, and $\gamma$ given now by
\[\alpha= 2q_1 q_6\,, \quad \beta=q_1 \overline{p_6} + q_6 \overline{p_1}\,,
\quad \gamma= 2(\overline{p_1p_6}-\overline{p_2p_5})\,.\] Formula
$(i)$ gives the second one of (\ref{ZCSmini}) and formula $(ii)$
gives (\ref{specialcon}).

Summarizing, $ZC$-states must be in one of the classes $ZCS$ and
$ZCE$ of the statement of the theorem. Viceversa, if a state is
$ZCS$, it is a separable state and therefore it is a $ZC$-state. If
a state is $ZCE$, it is straightforward to verify by plugging
(\ref{ZCEstati}) in the expressions (\ref{alfa}), (\ref{gamma}) and
(\ref{betaE}) that conditions $(i)$ and $(ii)$ of Lemma \ref{c1} are
verified for every concurrence. This concludes the proof of the
theorem. \qed

\section*{Appendix C: Proof of Lemma \ref{moreinfo}}

To simplify notations, it is convenient to use
$\alpha_{jk}:=(1-\lambda)p_j\overline{p_k}$ with $j \leq k$ and
$\beta_{jk}:=\lambda q_j q_k$. This way, $\rho^{T_A}$ writes as
\begin{equation}
\rho^{T_A}=\left(\begin{array}{cccccccc}\beta_{11}+\alpha_{11} &
\alpha_{12} & \alpha_{13}& 0 & \overline{\alpha_{15}} &
\overline{\alpha_{25}} & \overline{\alpha_{35}} & 0 \\
\overline{\alpha_{12}} & \alpha_{22} & \alpha_{23}& 0 &
\beta_{16}+\overline{\alpha_{16}} & \overline{\alpha_{26}} &
\overline{\alpha_{36}} & 0 \\ \overline{\alpha_{13}} &
\overline{\alpha_{23}} & \alpha_{33}& 0 & \overline{\alpha_{17}} &
\overline{\alpha_{27}} & \overline{\alpha_{37}} & 0 \\ 0 & 0 & 0 & 0
& \overline{\alpha_{18}} & \overline{\alpha_{28}} &
\overline{\alpha_{38}} & 0 \\ \alpha_{15} & \beta_{16}+\alpha_{16} &
\alpha_{17} & \alpha_{18} & \alpha_{55} & \alpha_{56} & \alpha_{57}
& \alpha_{58} \\ \alpha_{25} & \alpha_{26} & \alpha_{27} &
\alpha_{28} & \overline{\alpha_{56}} & \beta_{66}+ \alpha_{66} &
\alpha_{67} & \alpha_{68} \\ \alpha_{35} & \alpha_{36} & \alpha_{37}
& \alpha_{38} & \overline{\alpha_{57}} & \overline{\alpha_{67}} &
\alpha_{77} & \alpha_{78}
\\ 0 & 0 & 0 & 0 & \overline{\alpha_{58}} & \overline{\alpha_{68}}
& \overline{\alpha_{78}} & \alpha_{88}\end{array}\right).
\label{rhoTA}
\end{equation}

In our discussion,  we shall use the notation $PM(j_1,...,j_l)$ to
denote the principal minor  calculated as  the determinant of the
sub-matrix obtained by selecting the $(j_1,...,j_l)$ rows and
columns. For example $PM(1,2)$ denotes the  principal minor of order
$2$ obtained by calculating the determinant of the matrix at the
intersection of rows and columns 1 and 2. We shall use the Sylvester
criterion for a positive semi-definite matrix which says that an
Hermitian matrix is positive semi-definite if and only if all
principal minors
are nonnegative (see, e.g., \cite{Browne}, \cite{GantMA}).\\

Assume that $\rho$ is a PPT state.  By applying Sylvester criterion
with  $PM(4,5)$, $PM(4,6)$, $PM(4,7)$ in (\ref{rhoTA}), we obtain
that we must have $\alpha_{18}=\alpha_{28}=\alpha_{38}=0$. That is,
$p_8=0$ or $p_1=p_2=p_3=0$. However, if $p_1=p_2=p_3=0$,
$PM(2,5)=-\beta_{16}^2<0$, which is not possible. This establishes that $p_8=0$.\\

With this assumption, consider $PM(3,5,7)$ for (\ref{rhoTA}). A
direct calculation shows
\[PM(3,5,7)=\alpha_{77}\left(\overline{\alpha_{15}}
\alpha_{37}+\alpha_{15}\overline{\alpha_{37}}-\alpha_{55}
\alpha_{33}-
\alpha_{11}\alpha_{77}\right)=-(1-\lambda)^2|p_3p_5-\overline{p_1}\overline{p_7}|^2\,.\]
The last expression is positive only if
$p_3p_5=\overline{p_1}\overline{p_7}$. This implies
\begin{equation}
\alpha_{33} \alpha_{55}=\alpha_{11}\alpha_{77}\,. \label{TBU}
\end{equation}
We now show that (\ref{TBU}) cannot be with $\alpha_{77}\neq 0$,
therefore showing that $p_7$ must be zero. Assume that \eqref{TBU}
is true and $\alpha_{11}=0$. Then at least one between $\alpha_{55}$
and $\alpha_{33}$ must be zero. However $\alpha_{55}$ cannot be
zero, because this would give $PM(2,5)=-\beta_{16}^2<0$ and
$\alpha_{33}=0$ would require $PM(3,6)=-\alpha_{22}\alpha_{77}
\geqslant 0$, that is $\alpha_{22}=0$ which would lead again to
$PM(2,5)=-\beta_{16}^2<0$. Therefore, we must have $\alpha_{11} \neq
0$, which also, from orthogonality, implies $\alpha_{66}\neq 0$ and
from (\ref{TBU}) $\alpha_{33}\neq 0$ and $\alpha_{55} \neq 0$.
Moreover $\alpha_{22}\neq 0$ also is true  by considering  $PM(2,7)$
in \eqref{rhoTA}. Therefore, we are in the situation where {\it all}
the components of $\psi_{2}$, except $p_4$ and $p_8$, are different
from zero. Now, an argument as for $PM(3,5,7)$ above, applied this
time on $PM(2,3,6)$, along with the fact that $\alpha_{22}\neq 0$,
gives
\begin{equation}
\alpha_{66} \alpha_{33}=\alpha_{22}\alpha_{77}\,, \label{TBUbis}
\end{equation}
and
\begin{equation}
\alpha_{23}
\overline{\alpha_{67}}+\overline{\alpha_{23}}\alpha_{67}=\alpha_{22}\alpha_{77}+\alpha_{33}
\alpha_{66}=2\alpha_{33} \alpha_{66}\,. \label{usedlater}
\end{equation}
Combining \eqref{TBU} with \eqref{TBUbis}, we have
\begin{equation}
\alpha_{11}\alpha_{66}=\alpha_{22}\alpha_{55}\,. \label{TBUtris}
\end{equation}
We chose the overall  phase of $\psi^{(2)}$ such that
$q_1^2p_1^2=q_6^2p_6^2$ is real. Hence,
$p_1\overline{p_6}=\overline{p_1}p_6$, i.e.
$\overline{\alpha_{16}}=\alpha_{16}$. By multiplying
\eqref{usedlater} by $\alpha_{16}$, we obtain
\begin{equation}
\alpha_{23}\overline{\alpha_{17}}+\alpha_{17}
\overline{\alpha_{23}}=2 \alpha_{16} \alpha_{33}\,. \label{poi1}
\end{equation}
Calculation of $PM(2,3,5)$ gives, because of \eqref{TBU},
\[PM(2,3,5)= -\alpha_{22} \alpha_{33} \alpha_{55}+
(\beta_{16}+\alpha_{16}) \left( \alpha_{23}\overline{\alpha_{17}}+
\alpha_{17} \overline{\alpha_{23}}\right)- \alpha_{33}
\left(\beta_{16}+\alpha_{16} \right)^2\,.\] By replacing
\eqref{poi1} and using \eqref{TBUtris}, this expression simplifies
to
\[PM(2,3,5)=-\alpha_{33}\left(\alpha_{16}-(\beta_{16}+\alpha_{16})
\right)^2 =-\alpha_{33}\beta_{16}^2 <0\,.\]
This is not possible. Hence, \eqref{TBU} holds only if $p_7=0$.\\

Since $p_4=p_7=p_8=0$, consideration of $PM(2,7)$ and $PM(1,7)$ in
\eqref{rhoTA} shows that it must be $p_3=0$, or $p_6$ and $p_5$ both
equal to zero. However, the second case would imply
$PM(2,5)=-\beta_{16}^2 <0$. This establishes $p_3=0$ and concludes
the proof of the necessity of $p_3=p_4=p_7=p_8=0$. This shows that
if a state is $PPT$ its canonical form is written as
(\ref{formPPT}).

In order for $\rho$ to be a $PPT$-state the $4\times 4$ matrix
$\left(\begin{array}{cc}\rho_{11} & \rho_{12} \\  {\rho_{12}}^\dag &
\rho_{22}\end{array}\right)$ must be $PPT$ as a $2 \times 2$ state,
but since the $PPT$ test is necessary and sufficient for
separability in the $2\times 2$ case, this represents a $2 \times 2$
separable state. That is, there exist positive constants $\mu_j $,
$j=1,...,l$, with $\sum_{j=1}^l \mu_j=1$ and $2 \times 2$ density
matrices $\rho_j^{(1)}$, $\rho_j^{(2)}$ such that
\begin{equation}\label{poi2}\left(\begin{array}{cc}\rho_{11} & \rho_{12} \\
{\rho_{12}}^\dag & \rho_{22}\end{array}\right)=\sum_{j=1}^l \mu_j
\rho_j^{(1)} \otimes \rho_j^{(2)}\,. \end{equation} In particular,
\begin{equation}
\rho_{11}=\sum \mu_j \left(\rho_j^{(1)}\right)_{11}\rho_j^{(2)},
\quad \rho_{12}=\sum \mu_j \left({\rho_j^{(1)}}\right)_{12}
\rho_j^{(2)},\quad \rho_{22}=\sum \mu_j
\left({\rho_j^{(1)}}\right)_{22} \rho_j^{(2)}\,. \label{string}
\end{equation}
The $4\times 4$  matrices
\[\tilde \rho_j=\left(\begin{array}{cc}1 & 0 \\ 0 & 0\end{array}\right)\otimes \rho_j^{(2)}\,,\]
are density matrices and, using \eqref{string}, \eqref{poi2} and
(\ref{formPPT}), we obtain
\[\rho=\sum \mu_j \rho^{(1)}_j \otimes \tilde \rho_j\,,\]
which shows that $\rho$ is separable as well.

The fact that $\rho$ in the form (\ref{formPPT}) is a PPT-state
follows from the above characterization of $\rho$ as separable and
the
fact that every separable state is a PPT-state. \qed\\
\end{document}